\documentclass[prx,twocolumn,showpacs,preprintnumbers,amsmath,amssymb,superscriptaddress]{revtex4-1}

\usepackage{graphicx}
\usepackage{bm}

\begin{document}

\title{Bain distortion of noble metal thin films that exhibit fcc, bct, and reoriented fcc structures}

\author{Shota Ono}
\affiliation{Institute for Materials Research, Tohoku University, Sendai 980-8577, Japan}

\author{Koharu Tamura}
\affiliation{Department of Electrical, Electronic and Computer Engineering, Gifu University, Gifu 501-1193, Japan}

\begin{abstract}
A recent experiment has reported that body-centered cubic (bcc)-structured Ag is realized by bending face-centered cubic (fcc)-structured Ag nanowires [S. Sun {\it et al}., Phys. Rev. Lett. {\bf 128}, 015701 (2022)]. However, the bcc phase has been observed only near the Ag surface. Here, we explore how the bcc phase is stabilized near the surface by compressions. Our first-principles calculations for noble metals, Cu, Ag, and Au, indicate that body-centered tetragonal (bct) rather than bcc structure is preferred due to the surface effect. The bct-fcc boundary treated as a fixed boundary condition is necessary to thermodynamically stabilize the bct phase of Ag nanowire. The correlation between crystal structure and electron density-of-states is also discussed for three noble metals.
\end{abstract}

\maketitle

\section{Introduction} 
As the structure is correlated with the physical properties of solids, transformation pathways connecting different crystal structures have been explored for a long time, in particular for metallic systems \cite{bain,burgers,folkins,went,grimvall,togo,vdv}. They usually form one of three structures, {\it i.e.,} face-centered cubic (fcc), hexagonal close-packed (hcp), and body-centered cubic (bcc), at ambient conditions. The fcc-bcc and bcc-hcp transformations are known as tetragonal Bain and Burgers distortions, respectively \cite{bain,burgers}. Other related transformations have been studied analytically \cite{cayron2015,cayron2016}. In the Bain distortion, the ratio of lattice parameters characterizes the bcc ($c/a=1$) and fcc ($c/a=\sqrt{2}$) structures. Potential energy and free energy curves along the Bain path have been calculated by using first-principles approach for elemental metals and binary compounds \cite{grimvall,alippi,mehl,schonecker,wang,kolli,ono2021,ono_koba,wang2022,jerabek2022,navarro2024}. Most of calculations have shown an instability of fcc metals in the bcc structure. 

\begin{figure}
\center
\includegraphics[scale=0.47]{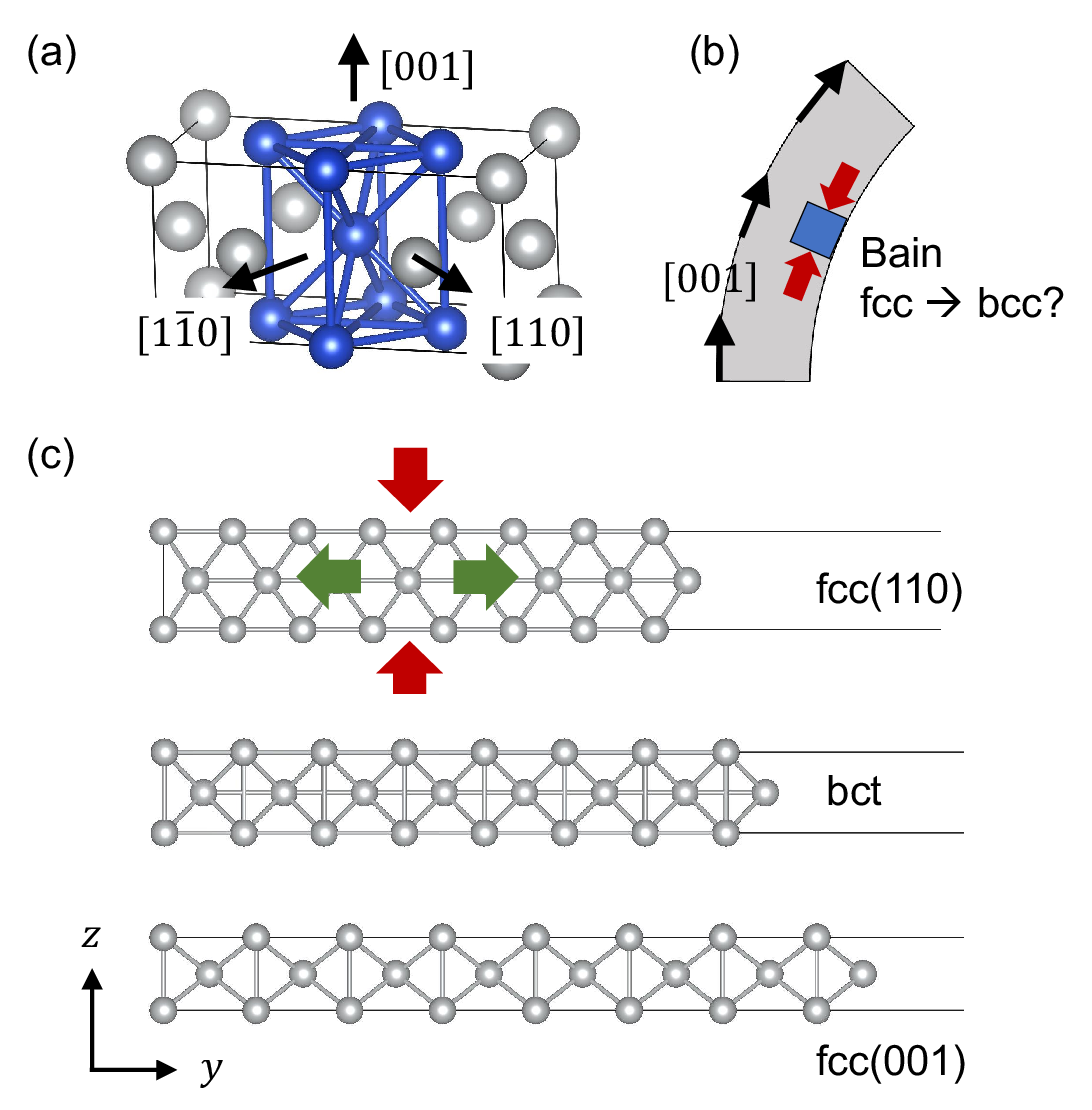}
\caption{(a) Tetragonal unit cell in the fcc structure. Atoms that form the tetragonal cell are highlighted by blue. Bulk Bain distortion, {\it i.e.}, a compression along the [001] direction, yields the bcc structure. (b) Schematic illustration of noble metal NW with the axis along the [001] direction. Bending the NW induces the Bain distortion around the surface. (c) The fcc thin film before and after the surface Bain distortion. The Bain distortion induces a compression and elongation along the [001] and [110] directions, respectively, and results in a bct structure followed by a reorientation of the fcc structure. }
\label{fig1} 
\end{figure}

Recently, Sun {\it et al}. have observed structural transformations of silver nanowires (Ag NW) that exhibit fcc to bcc to hcp to reoriented fcc structures under bending \cite{sun2022}. The Ag NW has the axis of [001] and the surface orientations of [110] and [$1\bar{1}0$] (see Fig.~\ref{fig1}(a)). Bending Ag NW creates a compressive force along the axis and around the (110) surface, which yields an fcc-bcc transformation. This has been interpreted as a tetragonal Bain distortion (see Fig.~\ref{fig1}(b)). However, bcc Ag has been observed only near the surface in experiment \cite{sun2022}. 
In addition, bcc-structured fcc metals (bulk phase) are known to be unstable at ambient condition \cite{grimvall}.
Therefore, the realization of bcc Ag is questionable. 

The transformation from fcc to bcc in Ag has been discussed in the field of high pressure physics \cite{sharma,coleman,bannon,smirnov}. Shock compression experiments have reported that bcc phase was observed around 150 GPa before an appearance of liquid phase at 180 GPa \cite{sharma,coleman}. Diamond anvil cell experiments and theoretical calculations have suggested that fcc Ag remains stable at 416 GPa and bcc Ag is lower in energy than fcc at 0.3 times the equilibrium volume \cite{bannon}. First-principles calculations have suggested that bcc phase is preferred at more than 100 GPa and 3000 K \cite{smirnov}. It is interesting if bcc phase is obtained by bending Ag NW without such high pressure and temperature conditions. 

In this paper, we study the Bain distortion of noble metals, Cu, Ag, and Au, by considering the surface effect. 
By using first-principles calculations, we demonstrate that the surface Bain distortion does not create a bcc structure but instead yields a body-centered tetragonal (bct) structure, which disagrees with experimental observations \cite{sun2022}. 
By performing finite-temperature molecular dynamics (MD) simulations, we show that the bct phase is thermodynamically stable only when a fixed boundary condition is imposed. This suggests that the bct phase should be supported by fcc phase that exhibits no structural transformations. 
We demonstrate that the electron density-of-states reflects the structural changes in the surface and bulk. 

Based on the semi-empirical force fields, Diao {\it et al.} \cite{diao} and Delogu \cite{delogu} have shown that the Ag and Au nanowires with cross-sectional area below 4 nm$^2$ contract spontaneously and undergoes an fcc-bct transition at zero pressure \cite{diao,delogu}. In contrast, the cross-sectional area of nanowires studied by Sun {\it et al.} \cite{sun2019} is large enough to keep the fcc phase.

A recent MD simulation has predicted a successive transformation from fcc to bcc to hcp structures in a Cu thin film when the system is stretched along the [100] direction \cite{sun2019}. However, periodic boundary conditions have been imposed to all directions in the simulated system, and no surfaces have been assumed. 

\begin{figure}
\center
\includegraphics[scale=0.45]{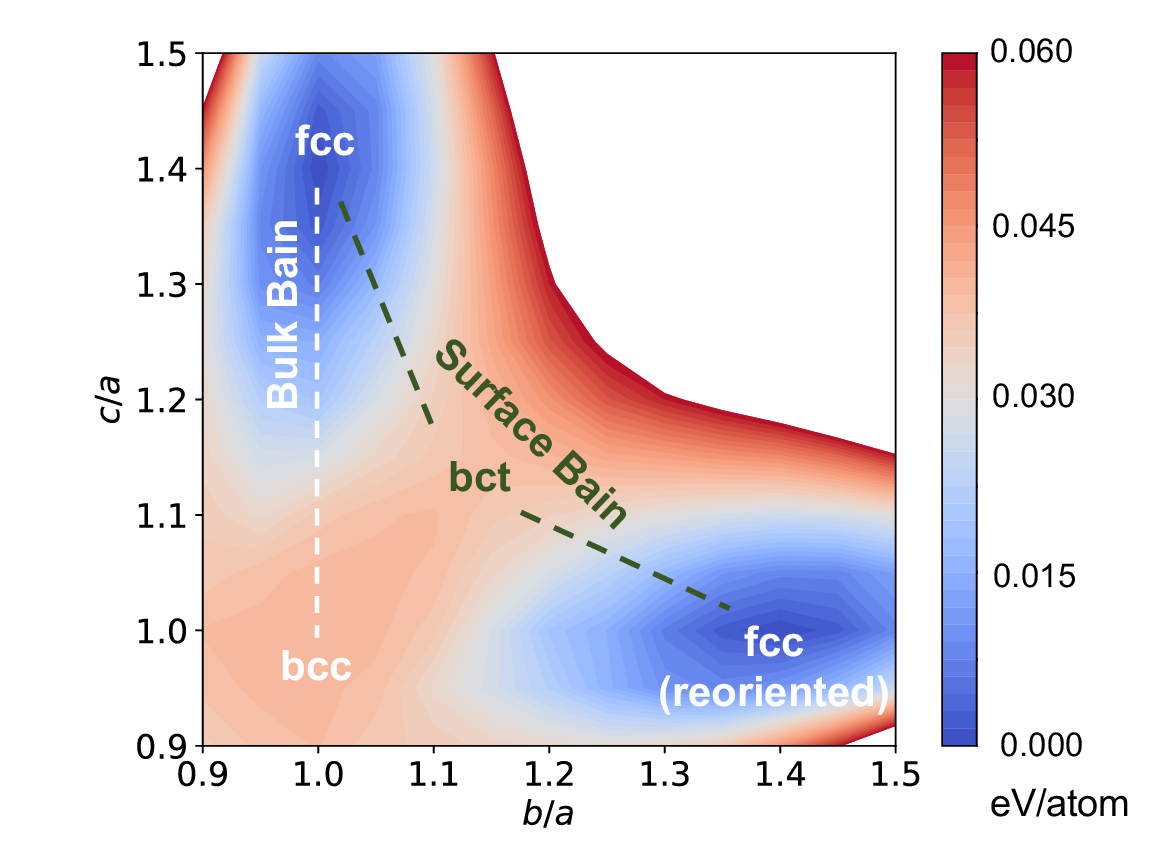}
\caption{Energy surface as functions of $c/a$ and $b/a$ in Ag bulk. The bulk Bain path connects fcc with bcc structure via tetragonal distortion by keeping $a=b$. The surface Bain path under free-boundary condition along $b$-axis direction connects fcc with reoriented fcc structure through bct that is lower than bcc structure in energy. }
\label{fig2} 
\end{figure}

\begin{figure*}
\center
\includegraphics[scale=0.45]{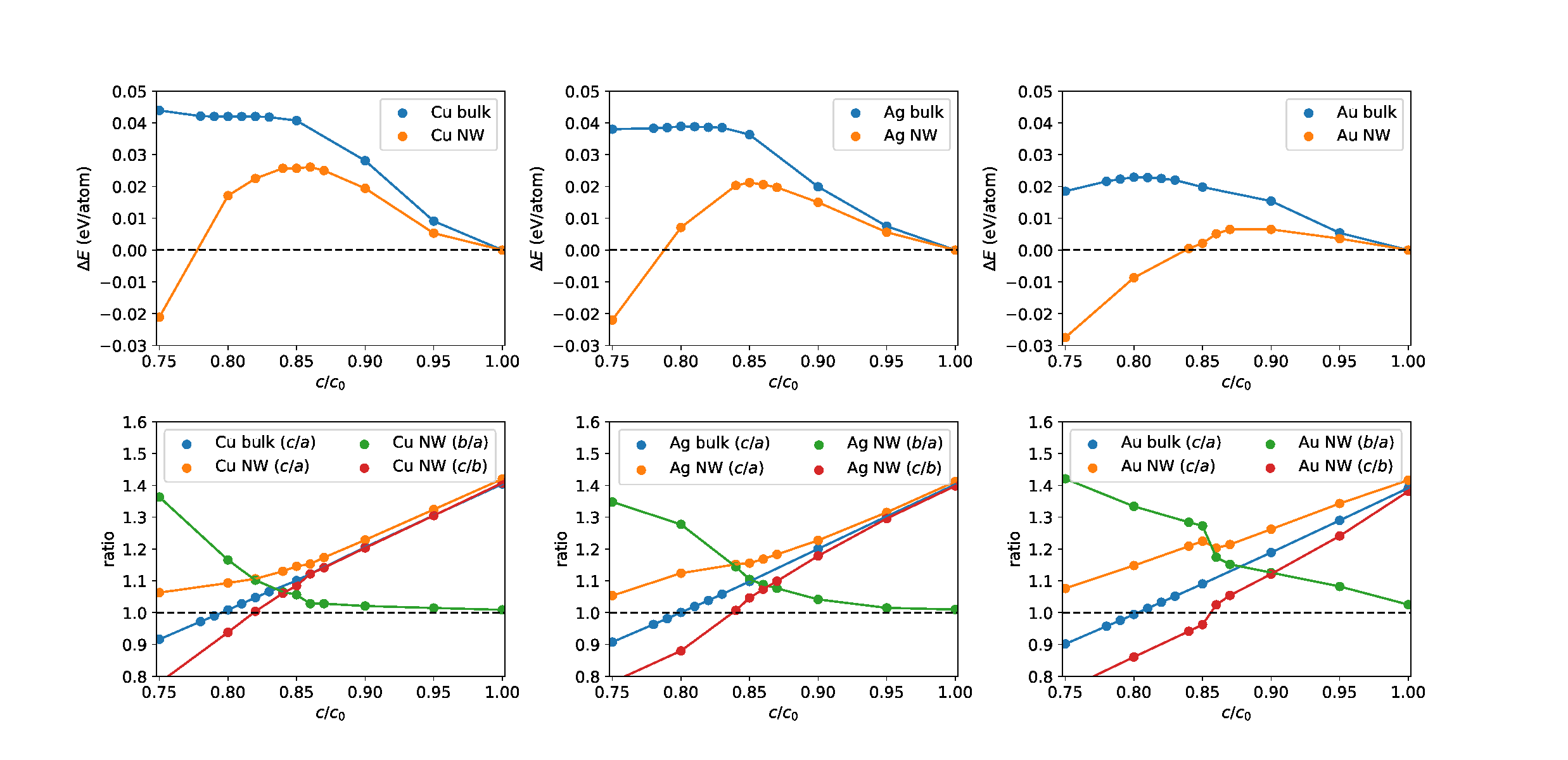}
\caption{Energy (upper) and the ratio of the lattice parameters ($c/a$, $b/a$, and $c/b$) along the bulk and surface Bain paths for Cu (left), Ag (middle), and Au (right) bulk and nanowire. The energy $\Delta E$ is measured from that without Bain distortion. The equality of $c/b=1$ corresponds to the bct structure. }
\label{fig3} 
\end{figure*}

\section{Bain distortion concept}
Bulk Bain distortion is a compression or elongation of tetragonal cell including two atoms. The ratio $c/a$ is exactly equal to $\sqrt{2}$ for the fcc structure (see Fig.~\ref{fig1}(a)). A compression along the [001] direction yields a decrease in $c$. To avoid a significant decrease in the unit cell volume, the tetragonal cell in turn expands along the lateral directions. When $c/a=1$, one obtains the bcc structure.  

We next consider a ``surface Bain distortion'' of fcc metal NWs. The NW has the axis parallel to the [001] direction and has the (110) and ($1\bar{1}0$) surfaces. We assume that the width along the [$1\bar{1}0$] direction is large enough to ignore the edge of metallic NW ($\sim 10$ nm in experiment \cite{sun2022}). As in the bulk Bain distortion, a compression along the [001] direction yields a lateral expansion to reduce strains stored in the volume. However, with the free-boundary condition along the [110] direction, atoms around the (110) surface can move largely to reduce the strain more efficiently. This produces a reoriented fcc structure in bulk, where the (110) surface is transformed into the (001) surface by the lateral compression (see Fig.~\ref{fig1}(c)). We investigate whether bcc-like phase appear along the distortion path. 

The physics underlying the reorientation is three-fold degeneracy of fcc structure.
When orthorhombic cell is used, there are three representations to express the fcc structure: $a=b=c/\sqrt{2}$ (Fig.~\ref{fig1}(a)), $a=b/\sqrt{2}=c$, and $a/\sqrt{2}=b=c$.
Figure \ref{fig2} shows a part of the potential energy surface of Ag bulk.
The volume of the unit cell containing two atoms is fixed to the equilibrium volume ($33.21$ \AA$^3$) of the fcc structure. 
Two energy minima correspond to the fcc and reoriented fcc structures, and an energy maximum corresponds to the bcc structure. 
Note also that bct structure is located at the saddle point. This is higher than fcc but lower than bcc structure by 6 meV/atom in energy. 
When the surface is present along the lateral direction, the constraint of $b/a=1$ is relaxed. 
Then, the surface Bain distortion will avoid the energy maximum of bcc phase.

In the rest of the paper, we address when the bct phase appears along the surface Bain distortion path, how such a metastable phase is stabilized at the ambient condition, and how the bct is distinguished from bcc phase, by using first-principles approaches. 
 
\section{Surface Bain distortion}
We first created a $1\times8\times1$ supercell (including 16 atoms) of Ag crystals with the [$1\bar{1}0$], [110], and [001] directions parallel to the $x$, $y$, and $z$ directions, respectively. We optimized the lattice parameters of the supercell and obtained $c_0=$ 4.03 \AA. We next prepared a 25 \AA-thick vacuum layer along the $y$ direction and shortened the lattice parameter $c$ along the $z$ direction. We then optimized the lattice parameter $a$ along the $x$ direction and relaxed all atomic positions in the slab model. Fixing $c$ corresponds to applying forces from bent nanowires. The optimized energy is a function of $c$, which is expressed as 
\begin{eqnarray}
 E(c) = \min_{a,\{ \bm{R}_i \}}E(a,c; \{ \bm{R}_i \}),
\end{eqnarray}
where $E(a,c; \{ \bm{R}_i \})$ is the total energy of the slab and $\{ \bm{R}_i \}$ is a set of positions of atom $i$.
This is similar to the uniaxial Bain path calculation for bulk crystals, where $c$ is fixed and $a$ is adjusted to minimize the total energy \cite{alippi}. The present calculation includes effects of the surface by considering $\{ \bm{R}_i \}$.  

Figure \ref{fig3} (upper middle) shows the energy variation as a function of $c/c_0$ for Ag. For bulk, $\Delta E$ takes a maximum value at $c/c_0=0.8$ (bcc structure). When metallic surface is laterally compressed, the increase in $\Delta E$ is moderate compared to the bulk case. $\Delta E$ takes a maximum value at $c/c_0=0.85$ and becomes negative for $c/c_0<0.8$. Negative value of $\Delta E$ indicates that energetically stable (001) surface appears and fcc structure is reoriented, as discussed below. The energy curve approaches the bulk case if the Ag film thickness is increased. 

Figure \ref{fig3} (lower middle) shows the structure parameter variations versus $c/c_0$ for Ag bulk and surfaces, where $a$ and $c$ are the lattice constant along the $x$ and $z$ directions, respectively. $b=d/n_y$ is the averaged lattice constant along the $y$ direction, where $d$ is the film thickness and $n_y (=7)$ is the number of unit cells. For bulk, $c/a$ decreases from $\sqrt{2}$ with decreasing $c/c_0$, and takes a unity at $c/c_0=0.8$. This corresponds to the bcc structure. For surface, $c/a$ is no longer a linear function of $c/c_0$. $b/a$ starts to increase around $c/c_0\simeq 0.85$ and reaches $\simeq \sqrt{2}$. In this way, the surface Bain distortion results in a reorientation of fcc structure. 

What is important is that the crystal structure of Ag NW at $c/c_0=0.84$ is almost tetragonal (see the curve of $c/b$ (red)). This corresponds to $a=2.94$ \AA, $b=3.36$ \AA, and $c=3.38$ \AA, which can be seen as a square lattice when viewed from the $a$-axis direction. Interestingly, the lattice constants of $b$ and $c$ are quite close to the experimental value (2.36$\times \sqrt{2}\simeq$3.33 \AA) \cite{sun2022}. The latter has been estimated from the distance between \{110\} planes, while the length of $a$ has not been measured. The present calculations suggest that bct rather than bcc phase is realized in Ag NW, which is due to the surface effect that induces asymmetry of $x$ and $y$ directions, as discussed in Fig.~\ref{fig2}. The structural parameters are listed in Table \ref{table1}.

We have also studied the surface Bain distortion for Cu and Au, as shown in Fig.~\ref{fig3} (left and right).
The optimized values of $c_0$ are 3.56 and 4.01 \AA \ for Cu and Au, respectively.
The $c/c_0$-dependences of $\Delta E$ and structural parameter ratio are similar to those of Ag case, whereas the maximum of $\Delta E$ in Au is smaller than in Cu and Ag by a factor of two.
The fcc-bct-fcc transformation is also observed, while the value of $c/c_0$ at which the bct phase appears is different from Ag: $c/c_0\simeq 0.82$ and 0.86 for Cu and Au, respectively. For Au NW, $c/a, b/a$, and $c/b$ exhibit large changes around $c/c_0\simeq 0.86$.

\begin{table}\begin{center}\caption{
Lattice parameters in unit of \AA \ for Ag NW and bulk systems. 
}
{\begin{tabular}{lccc}\hline\hline
 \hspace{4mm} & $a$ \hspace{4mm} & $b$ \hspace{4mm} & $c$ \hspace{4mm}\\ \hline
Ag NW ($c/c_0=1$) \hspace{4mm} & 2.85 \hspace{4mm} & 2.85 \hspace{4mm} & 4.03 \hspace{4mm}\\
Ag NW ($c/c_0=0.84$) \hspace{4mm} & 2.94 \hspace{4mm} & 3.36 \hspace{4mm} & 3.38 \hspace{4mm}\\
bcc Ag \hspace{4mm} & 3.23 \hspace{4mm} & 3.23 \hspace{4mm} & 3.23 \hspace{4mm}\\
bct Ag \hspace{4mm} & 3.03 \hspace{4mm} & 3.33 \hspace{4mm} & 3.33 \hspace{4mm}\\
experiment \cite{sun2022} \hspace{4mm} & - \hspace{4mm} & 3.33 \hspace{4mm} & 3.33 \hspace{4mm}\\
\hline
\end{tabular}
}
\label{table1}\end{center}\end{table}

\begin{figure}
\center
\includegraphics[scale=0.6]{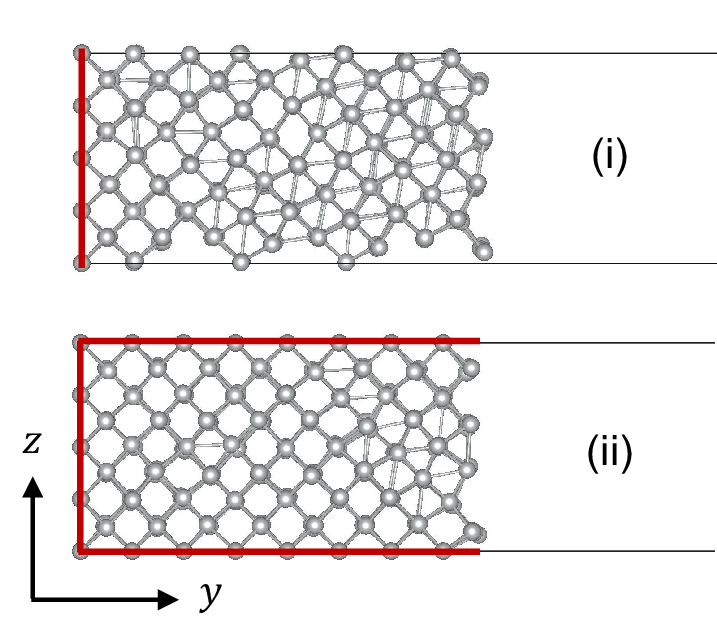}
\caption{Atomic distribution of Ag NW at $c/c_0=0.84$ after a MD simulation (1.5 ps for 300 K). Ag atoms on the plane (solid red), $y=0$ for model (i) and $y=0$ and $z=0$ for model (ii), are fixed during the simulation. }
\label{fig4} 
\end{figure}

The stability of the bct phase against the thermal fluctuations was investigated by performing finite-temperature MD simulation. We created a $4\times 4$ supercell (256 atoms), where the unit cell of the optimized Ag NW at $c/c_0=0.84$ is repeated along the $x$ and $z$ directions. To stabilize the bct phase, we considered two fixed boundary conditions: (i) atoms on the $y=0$ plane are fixed and (ii) atoms on the $y=0$ and $z=0$ planes are fixed. Figure \ref{fig4} shows an atomic configuration at 1.5 ps for 300 K. For model (i), the bct phase is disordered over the system. For model (ii), such a disordering is weak enough not to cause any phase transformations except for near the surface. In our MD simulation, the fixed boundary condition plays a role of the bct-fcc boundary. In experiment \cite{sun2022}, the bcc-like phase is in fact surrounded by fcc ground state. Our simulation implies that the bct phase is stabilized by the fcc phase that exhibits no structural disorders.

\begin{figure*}
\center
\includegraphics[scale=0.42]{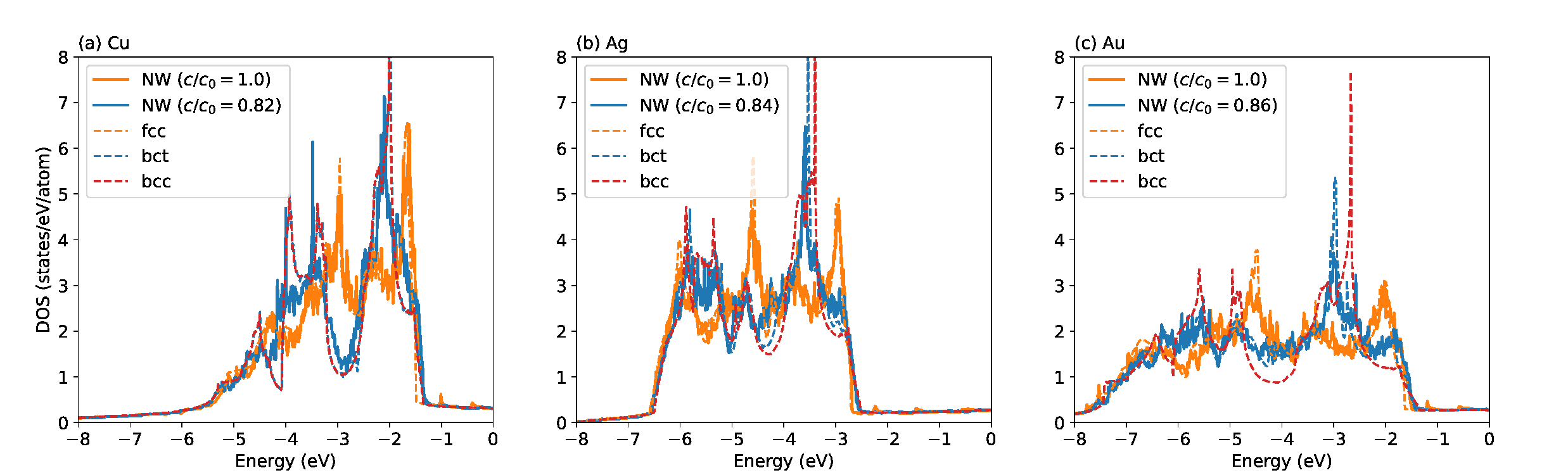}
\caption{Electron DOS per atom for noble metal NWs at $c/c_0=1.0$ and 0.82 (Cu), 0.84 (Ag), and 0.86 (Au), and for the bulk in the fcc, bct, and bcc structures (dashed). The energy is measured from the Fermi level. The curves of bct Cu and bcc Cu overlap. }
\label{fig5} 
\end{figure*}

Figure \ref{fig5}(b) shows the electron density-of-states (DOS) for Ag NW at $c/c_0=1.0$ and 0.84. 
For comparison, the DOS of Ag bulk in the fcc, bct, and bcc structures are also shown. 
The lattice constants of the bct structure are similar to the structural parameters of Ag NW at $c/c_0=0.84$, as listed in Table \ref{table1}. 
Below the Fermi level, the contribution from the $4d$ bands ranges from $-2.75$ to $-6.5$ eV when $c/c_0=1.0$. 
When the surface Bain distortion is applied ($c/c_0=0.84$), the higher band edge is smeared out and the DOS starts to increase at $-2.5$ eV.
In addition, the DOS peak at $-3.0$ eV is broadened and a strong peak appears at $-3.6$ eV.
The DOS of fcc and bct Ag (bulk) are similar to those of Ag NW at $c/c_0=1.0$ and 0.84, respectively. 
For bcc Ag (bulk), DOS shows a strong peak at $-3.4$ eV. 
This is followed by a nearly flat DOS ($-3.4\sim -3.75$ eV), which is absent in bct Ag. 
Such a difference will be helpful to distinguish the bct Ag from the bcc Ag in future experiments.

The electron DOS for Cu and Au systems are also shown in Fig.~\ref{fig5}(a) and \ref{fig5}(c), respectively. 
The Cu-$3d$ and Au-$5d$ bands start to contribute to the DOS below $-1.5$ eV.
Basically, the structure dependence of the DOS is similar to the Ag system.
In fact, when $c/c_0$ is decreased, a strong peak around the higher band edge shifts to lower energy. 
A nearly flat DOS is observed around $-3$ eV for bcc Au but is absent for bct Au. However, for Cu bulk, the bct structure relaxes to the bcc structure, and their DOS curves overlap.

\section{The bct-hcp transformation}
The Burgers distortion relates the bcc with hcp structures \cite{burgers}. 
This is done by shifting the lattice plane alternately along the [110] direction of bcc structure and adjusting the lattice constants to form a triangular lattice on the (110) plane. The [110] direction is then regarded as the [0001] direction in the hcp structure. 
Sun {\it et al.} have explained the bcc-hcp transformation of Ag NW based on the Burgers distortion \cite{sun2022}. 
This is naturally extended to the bct-hcp transformation by considering the shortest edge of the tetragonal cell as the $c$ axis. 
Note that the lattice constant of hcp structure is $2.87$ \AA, which is comparable with $3.03$ \AA \ of the bct structure and is much smaller than $3.23$ \AA \ of the bcc structure (see Table \ref{table1}). 

\section{Conclusion}
By using first-principles calculations, we have studied the Bain distortion of the noble metal NW with surfaces and investigated whether bcc structure is realized by compressions. In contrast to the experimental observations for Ag NW \cite{sun2022}, we have not obtained bcc structure but bct structure. Our MD simulation for Ag NW has suggested that the bct phase should be surrounded by fcc phase to avoid structural transformations. We have demonstrated that the electron DOS reflects the structural changes in the surface and bulk. 

The present work has extended the Bain distortion concept to apply the surface systems and indicated that the surface (without periodic boundary conditions) could play an important role in structural transformations. In this sense, bending nanowires having a specific surface would produce metastable phases without high pressure and temperature conditions \cite{sharma,coleman,bannon,smirnov}. However, the metastable phase created well below the surface has to be supported by ground state phase. It is interesting to explore electronic and magnetic properties that are intrinsic to metastable structures of other metallic systems. 

\section*{Appendix: Computational details}
We performed first-principles calculations by using \texttt{Quantum ESPRESSO} \cite{qe} based on density-functional theory. The ultrasoft pseudopotentials in \texttt{pslibrary.1.0.0} \cite{dalcorso} was used to treat the electron-ion interaction. The modified Perdew-Burke-Ernzerhof generalized gradient approximation (PBEsol-GGA) was adapted for the exchange-correlation energy \cite{pbesol}. For the case of gold, we used the Perdew-Zunger local-density approximation \cite{pz}. These choices of exchange-correlation functionals are suitable for studying high pressure condition \cite{pbesol_metals}. The wavefunction (charge density) cutoff energy was set to be 60 (600) Ry. The $k$-points distance was set to be less than 0.15 \AA$^{-1}$ and the smearing parameter of 0.01 Ry \cite{smearingMP} was used in the self-consistent field (scf) calculations. Electron DOS was calculated by using tetrahedron method \cite{bloechl}. In the MD simulations, the volume of the supercell was fixed and the $\Gamma$ point sampling was used. The Newton's equation was integrated using the Verlet algorithm with a time step of 2 fs. The ionic temperature was controlled using the velocity scaling and kept to $T=$ 300 K. 

We used \texttt{Atomic Simulation Environment} \cite{ase} for modeling the metallic NWs and also used \texttt{VESTA} to visualize crystal structures \cite{vesta}.

\begin{acknowledgments}
We thank T. Nitta for useful discussions. This work was supported by JSPS KAKENHI (Grant No. 21K04628). A part of numerical calculations has been done using the facilities of the Supercomputer Center, the Institute for Solid State Physics, the University of Tokyo, and the supercomputer ``MASAMUNE-IMR'' at Center for Computational Materials Science, Institute for Materials Research, Tohoku University.
\end{acknowledgments}



\end{document}